# Excited states in the full QCD hadron spectrum on a $16^3 \times 40$ lattice

Dong Chen[a] [*]

[a]Department of Physics, Columbia University, New York, NY 10027, USA

We report the hadron mass spectrum obtained on a $16^3 \times 40$ lattice at $\beta = 5.7$ using two flavors of staggered fermions with $ma = 0.01$. We calculate the masses of excited states that have the same quantum numbers as the $\pi$, $\rho$ and $N$. They are obtained by a combined analysis of the hadron correlators from sources of size $16^3$ and $8^3$. We also report on the hadron spectrum for a wide range of valence quark masses.

## 1. INTRODUCTION

The hadron mass spectrum is of fundamental interest in lattice QCD. The Columbia group has done hadron mass calculations with two flavors of staggered dynamical quarks for $ma = 0.01, 0.015, 0.02, 0.025$ on a $16^3 \times 32$ volume and $ma = 0.01, 0.025$ on $32^4$, both at $\beta = 5.7$ [1–3]. We have also investigated $ma = 0.004$ on a $16^3 \times 32$ volume at $\beta = 5.48$ [4].

At Lattice '94, we reported a new study on a $16^3 \times 40$ lattice with $ma = 0.01$ at $\beta = 5.7$ [5]. Since then, we have extended our calculation to 4900 micro-canonical time units. We systematically study the effects of source size on the determination of the mass values. We obtain the excited states by combined analysis of hadron propagators from different sized sources. We also report on the hadron mass spectrum with a variety of valence quark masses and compare our results with our earlier runs.

## 2. SIMULATION

Table 1 lists the parameters of our calculation. We used the 'R'-algorithm of Gottlieb, *et. al.* [6] for our evolution, along with their notation. We have collected 4900 micro-canonical time units during a 7.5 month run on the 256-node Columbia parallel computer. Hadron masses were measured every 6 time units. For a single gauge configuration, we did 5 measurements with the source set on 5 different time slices and averaged the result over time slices (AOTS).

Table 1
Simulation parameters

| | |
|---|---|
| volume | $16^3 \times 40$ |
| $\beta$ | 5.7 |
| $m_{dynamical}a$ | 0.01 |
| length | 250+4650 |
| trajectory length | 0.5 |
| step size | 0.0078125 |
| CG stopping condition | $1.13 \times 10^{-5}$ |
| avg CG steps | 660 |
| minutes/trajectory | 12.5 |
| hadron measured every | 6 |
| hadron source types | wall, 2Z |
| | wall, Z |
| hadron source sizes | $16^3, 12^3, 8^3, 4^3$, point |
| valence quark masses | 0.004, 0.01, 0.015, 0.02 |
| | 0.025, 0.05, 0.07 |

In our hadron mass calculations, the 2Z source is set on a fixed $t$ slice and is non-zero when only all $(x, y, z)$ are even; our Z source is non-zero for all $(x, y, z)$. We set our 2Z and Z wall sources across the entire spatial volume ($16^3$) as well as on smaller cubic spatial volumes ($12^3, 8^3, \ldots$). We also calculate quark propagators using different valence quark masses, as listed in Table 1. Out of all the possible combinations of source types, sizes and valence quark masses, we have a total of 19 combinations for mesons, 21 combinations for the nucleon and 10 combinations for the $\Delta$.

[*]This work was done in collaboration with Shailesh Chandrasekharan, Norman H. Christ, Weonjong Lee, Robert D. Mawhinney and Decai Zhu and was supported in part by the Department of Energy.

Figure 1 shows the evolution of $\langle \overline{\chi}\chi \rangle$ during our run. Discarding the first 250 time units for thermalization gives $\langle \overline{\chi}\chi \rangle = 0.0274(2)$. This is consistent with the value $0.0277(3)$ from our earlier $16^3 \times 32$ run [1,2].

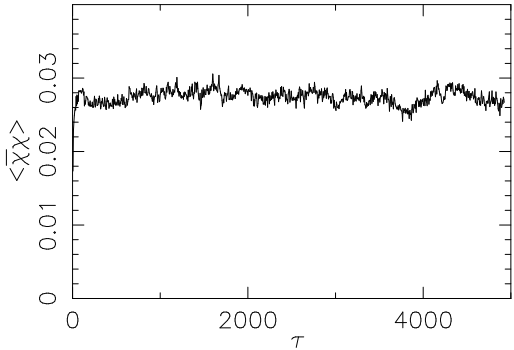

Figure 1. The evolution of the chiral condensate.

## 3. RESULTS

We reported source size effects on the effective mass plots for hadrons in ref [5]. Here we reproduce two such plots with more statistics. Figures 2 and 3 show from bottom to top the effective mass plots of the $\pi, \rho$ and $N$ for $16^3$ and $8^3$ sources. These figures demonstrate that $16^3$ is the optimal source size for this calculation. However, they also show that close to the source, there are signals for excited states since the effective masses are heavy.

Since the excited states have the same quantum numbers as the lowest states, a straight-forward fit to both lowest and excited states tends to have oscillations between the mass parameters resulting in poorly determined masses. We thus make combined fits using both $16^3$ and $8^3$ sources. The $16^3$ data are only fitted to the lowest states while the $8^3$ data are fitted to both lowest and excited states with the lowest states mass set identical to those for the $16^3$ data. (Each particle comes with its opposite parity state for staggered fermions. The $\pi$'s parity partner does not exist in the numerical data.) As an example, for the pion we

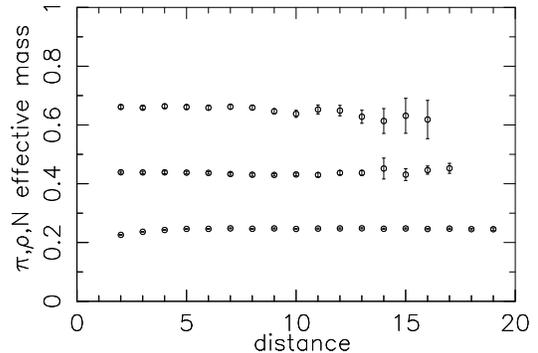

Figure 2. $\pi, \rho, N$ effective masses for $16^3$ source.

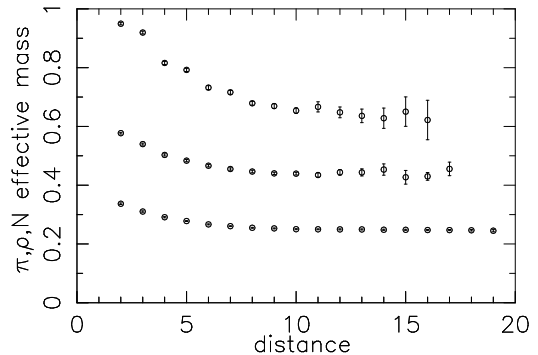

Figure 3. $\pi, \rho, N$ effective masses for $8^3$ source.

perform a one state fit to the $16^3$ source with

$$P_\pi(t) = A_0 \cosh(m_0 (t - N_t/2)), \qquad (1)$$

and a two state fit to the $8^3$ source with

$$P_\pi(t) = A'_0 \cosh(m_0 (t - N_t/2)) \\ + A'_2 \cosh(m'_2 (t - N_t/2)). \qquad (2)$$

The two $m_0$'s are set to be the same; thus the combined fit has $2 + 4 - 1 = 5$ parameters. For all other particles, we make a 2 state fit to the $16^3$ source and a 4 state fit to the $8^3$ source with the two lowest mass values identical and hence the combined fit has $4 + 8 - 2 = 10$ parameters. The degree of freedom ($dof$) for our fits is thus $2 \times (N_t/2 - t_{min})$ minus the number of fitting parameters. We include the full correlation matrix in all our fits.

Figures 4 and 5 show the combined fitting results for the $\pi$ and $\rho$. These are not effective mass plots, but the fitted masses from a distance $t_{min}$ up to $N_t/2$ as a function of $t_{min}$. There are clear signals for excited states at small $t_{min}$.

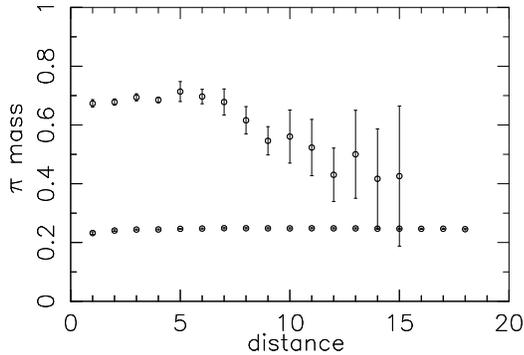

Figure 4. The $\pi$ and its excited state from a combined fit for the $16^3$ and $8^3$ sources.

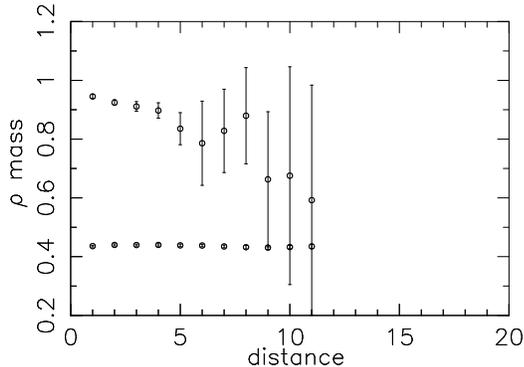

Figure 5. The $\rho$ and its excited state from a combined fit for the $16^3$ and $8^3$ sources.

Table 2 is a list of hadron masses for our current $16^3 \times 40$ calculation, as a function of the valence quark masses we used to calculate our hadron propagators. All our fits are correlated fits from $t_{min}$ to $N_t/2$ with errors determined from the jackknife method. The pion mass is obtained from a single state fit with the 2Z source. All other mesons and the nucleon come from two state fits with the 2Z source. Our $\Delta$'s come from two state fits with the Z source. $\Delta_1$ and $\Delta_0$ are two equivalent lattice operators corresponding to the two $A_2$ operators in ref [7], respectively. All our sinks are local except for the $\Delta$'s which are necessarily non-local in the staggered fermion formalism. All our source and sink time slices are fixed to Coulomb gauge. For the $\Delta$, we also measure in Landau gauge for comparison.

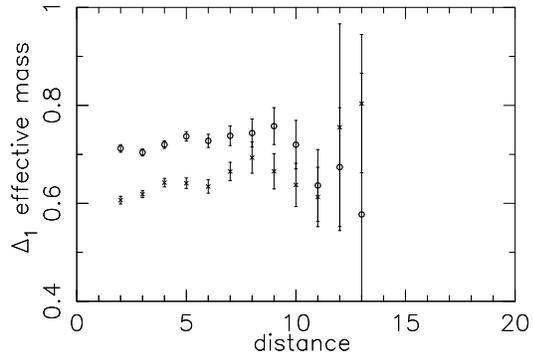

Figure 6. The $\Delta_1$ effective masses in Coulomb gauge ($\circ$) and Landau gauge ($\times$) for a $16^3$ source.

Figure 6 is an effective mass plot for the $\Delta_1$. The $\Delta_0$ plot is very similar. For small $t$, the $\Delta$ masses measured in Landau gauge are quite different from those measured in Coulomb gauge. They are about 3 standard deviations lower and are in closer agreement with the nucleon. This raises a question about the validity of using Landau gauge, non-local operators in general, since the gauge-dependent correlation between different time slices in Landau gauge may alter the signal one wants to obtain.

Table 3 lists the excited states in our calculation. We only measured $8^3$ source data for $ma = 0.01$. The $t_{min}$ used is less than in table 2 so that we have a better estimate for the excited states. Were we to use the same $t_{min}$ in table 2 as in table 3, we would have $< 1\%$ change in the mass values but a larger $\chi^2$. The lowest states from our combined $16^3$ source and $8^3$ source fits also agree with those in table 2 to 1%. All our fits have $\chi^2/dof \leq 2$. There is no apparent excited state for the $\pi_2$ in our data.

Compared with our earlier $16^3 \times 32$ calculation [2], where we have $m_\pi a = 0.252(3)$, $m_\rho a = 0.454(4)$ and $m_N a = 0.692(6)$, our current $16^3 \times 40$ mass spectrum is lighter. The nucleon is especially light, about 3 standard deviations lower than the earlier result. In our current calculation,

Table 2
Hadron mass results: $16^3 \times 40, \beta = 5.7, ma = 0.01$

| $m_{val}a$ |  | 0.004 | 0.01 | 0.015 | 0.02 | 0.025 | 0.05 |  | 0.07 |
|---|---|---|---|---|---|---|---|---|---|
|  | $t_{min}$ |  |  |  |  |  |  | $t_{min}$ |  |
| $m_\pi a$ | 11 | 0.167(2) | 0.248(2) | 0.299(2) | 0.343(1) | 0.382(1) |  | 15 | 0.643(1) |
| $m_{\pi_2} a$ | 8 | 0.210(7) | 0.285(3) | 0.335(3) | 0.380(3) | 0.421(3) |  | 11 | 0.702(2) |
| $m_{f_0} a$ |  | 0.320(13) | 0.424(11) | 0.474(10) | 0.516(9) | 0.553(9) |  |  | 0.811(4) |
| $m_\rho a$ | 6 | 0.407(5) | 0.438(4) | 0.465(4) | 0.492(4) | 0.519(4) |  | 12 | 0.754(3) |
| $m_{b_1} a$ |  | 0.590(36) | 0.599(11) | 0.626(9) | 0.654(9) | 0.683(9) |  |  | 0.940(29) |
| $m_{\rho_2} a$ | 6 | 0.398(10) | 0.433(6) | 0.463(5) | 0.492(4) | 0.521(4) |  | 10 | 0.758(3) |
| $m_{a_1} a$ |  | 0.513(13) | 0.564(11) | 0.598(11) | 0.629(10) | 0.660(10) |  |  | 0.905(9) |
| $m_N a$ | 6 | 0.600(10) | 0.661(7) | 0.708(6) | 0.753(6) | 0.797(5) |  | 11 | 1.163(5) |
| $m_{N'} a$ |  | 0.674(18) | 0.757(12) | 0.815(11) | 0.866(12) | 0.914(11) |  |  | 1.256(13) |
| $m_{\Delta_0} a$ | 6 | 0.734(33) | 0.746(16) |  |  | 0.852(8) | 1.040(6) |  |  |
| $m_{\Delta_1} a$ | 6 | 0.752(47) | 0.729(14) |  |  | 0.843(9) | 1.035(6) |  |  |

Table 3
Excited states, $m_{val}a = m_{sea}a = 0.01$

|  |  | $t_{min}$ | $\chi^2$ |
|---|---|---|---|
| $m_\pi a$ | 0.68(4) | 7 | 49(17) |
| $m_{\pi_2} a$ |  |  |  |
| $m_{f_0} a$ | 0.73(6) | 6 | 18(10) |
| $m_\rho a$ | 0.90(3) | 4 | 36(19) |
| $m_{b_1} a$ | 0.98(12) |  |  |
| $m_{\rho_2} a$ | 0.91(7) | 4 | 40(16) |
| $m_{a_1} a$ | 0.94(5) |  |  |
| $m_N a$ | 1.17(2) | 4 | 32(17) |
| $m_{N'} a$ | 1.19(3) |  |  |

The excited states are labeled as their corresponding lowest states. Other particles that have different quantum numbers in the continuum can also have projections onto these states on the lattice.

the masses do not depend on the upper bound of the fitting range within errors. The length in micro-canonical time units where we do hadron measurements for the new run is about 3 times longer than the old one, although the old run has more averages for a single gauge configuration. We would argue that in our previous run, the errors are underestimated because our simulation time is not sufficiently large to accurately compute the effects of autocorrelations.

Using the $\rho$ mass of 770 MeV to fix the scale, we find at $ma = 0.01$ that $m_\pi = 436(4)$ MeV, $m_N = 1162(12)$ MeV and $m_\Delta = 1297(28)$ MeV (averaging over our Coulomb gauge $\Delta_0$ and $\Delta_1$). The excited states have $m_\pi = 1195(70)$ MeV, $m_\rho = 1582(53)$ MeV and $m_N = 2057(35)$ MeV. Our lattice spacing is $a^{-1} = 1757(16)$ MeV and the lattice size is $La = 1.80(2)$ fm.